# Two-dimensional (2D) *d*-Silicates from abundant natural minerals

*Preeti Lata Mahapatra[a], Appu Kumar Singh[b], Raphael Tromer[c], Karthik R.[b], Ambresha M.[d], Gelu Costin[e], Basudev Lahiri[f], Tarun Kumar Kundu[b], P. M. Ajayan[g], Douglas S. Galvao*[c], Chandra Shekhar Tiwary*[b]*

[a]School of NanoScience and Technology, Indian Institute of Technology, Kharagpur, West Bengal-721302 India

[b]Metallurgical and Materials Engineering, Indian Institute of Technology Kharagpur, Kharagpur 721302, India

[c]Applied Physics Department, State University of Campinas, Campinas, SP, 13083-970, Brazil

[d]Materials Engineering, Indian Institute of Science, Bangalore, India-560012

[e]Department of Earth, Environmental and Planetary Sciences, Rice University, Houston, Texas 77005, United States

[f]Department of Electronics and Electrical Communication Engineering, Indian Institute of Technology Kharagpur, Kharagpur 721302, India

[g]Department of Materials Science and Nanoengineering, Rice University, 6100 S Main Street, Houston, TX, 77005, USA

[*]Corresponding author E-mail: chandra.tiwary@metal.iitkgp.ac.in, galvao@ifi.unicamp.br

**Abstract**

In the last decade, the materials community has been exploring new 2D materials (graphene, metallene, TMDs, TMCs, MXene, among others) that have unique physical and chemical properties. Recently, a new family of 2D materials, the so-called 2D silicates, have been proposed. They are predicted to exhibit exciting properties (such as high catalytic activity, piezoelectricity, and 2D magnetism). In the current work, we demonstrate a generic approach to the synthesis of large-scale 2D silicates from selected minerals, such as Diopside (*d*). Different experimental techniques were used to confirm the existence of the 2D structures (named 2D-*d*-silicates). DFT simulations were also used to gain insight into the structural features and energy harvesting mechanisms (flexoelectric response generating voltage up to 10 V). The current approach is completely general and can be utilized for large-scale synthesis of 2D silicates and their derivatives, whose large-scale syntheses have been elusive.

# 1. Introduction

The journey of two-dimensional (2D) materials started from the excitement of the possibility to separate single layers of carbon atoms from their bulk form of graphite. Continuing such material exploration, the research community demonstrated the synthesis of a single element such as metallene[1] (atomically thin tin, gallium, etc)[2], silicene (2D-Si)[3], germanene (2D-Ge)[4], phosphorene (2D-Phosphorus)[5], Borophene (2D-Boron)[6], etc. The synthesis of 2D-MoS2 from its naturally occurring bulk counterpart resulted in a new class of material, which is known as transition metal di-chalcogenides (TMDCs)[7]. These materials consisting of two elements (Mo (metal) and S) provide the opportunity to engineer them depending on Mo/S ratio. Both graphene and MoS2 have intrinsic layered structures that favor 2D morphology. This was further extended into alloy 2D, which includes groups of selenides, tellurides[8], and so on. TMDCs introduced direct bandgap in their monolayers, which led to an evolution in high-quality field effect transistors. Alloy 2D, on the other hand, was initially difficult to achieve due to its non-directional bonding, and yet researchers have discovered methods to create it. Due to their strong light-matter interactions, these 2D alloys have significant potential for developing superior optoelectronic devices[9].

Another class of functional materials is 2D metal oxides with complex composition and charge density are just a few of the physicochemical properties that can be easily changed in these structures. Recent advances in the synthesis of multifaceted two-dimensional (2D) materials provide opportunities and a dynamic platform for studying manifestations in fundamental research, such as superconductors, ferromagnetism, massless Dirac fermions, and quantum Hall effect[10]. To date, several categories of oxides have emerged in their 2D state, including metal mono-oxides (2D NiO, ZnO, CuO, etc.), metal di-oxides (2D $MnO_2$, $TiO_2$, $SnO_2$, and others), and complex metal oxides (2D perovskite, spinel, garnet, wurtzite, etc.)[10]. The delamination of non-layered complex metal oxides has recently gained much attention.

However, exfoliated non-layered structures could be slightly non-uniform in shape in contrast to layered materials. The non-layered material exfoliation results in abundant dangling bonds and defects on the surface with large crystal distortions[10]. A major class of such complex natural oxides is silicates, such as 2D mica[11,12], muscovite[13], biotene[14], and so on. Recently, Eric *et al.* predicted and demonstrated the formation of 2D transition metal silicates in a bottom-up approach[15]. The study observed that the interaction of the metal substrate and the grown silica layer is strong, making the exfoliation difficult. In order to have effective exfoliation, several approaches have been tried to confirm the reaction and weaken the metal-silicate and substrate site bonding via hydrogenation[15–17] Al-doping, water chemistry[18,19], and Molecular Beam Epitaxial (MBE) method[20,21].

A similar natural silicate mineral found abundantly on the earth's crust in parts of metamorphic rocks is a monoclinic pyroxene mineral called diopside with the general formula "CaMgSi2O6". It falls under the category of single chain inosilicate minerals with interlocking chains of silicate tetrahedra found in many igneous and metamorphic rocks (ultramafic). Diopside (*d*-silicate) has good mechanical properties, including a bending strength of 300MPa and a fracture toughness of 3.5 Mpa m1/2. It is also a biocompatible material and a good candidate for orthopedic coatings. Diopside ceramics promote apatite formation and form strong bonds with bone tissue. Under pressure, bones generate electricity, and this electromechanical behavior is considered to be critical for bone self-repair and remodeling properties. Diopside has the monoclinic structure and C2/C (C2H-6) space group with an inversion center. The bulk unit cell consists of eight Si-O tetrahedra, six Mg-O octahedra, and six Ca-O octahedra. In the current paper, we demonstrate an easily scalable and simple synthesis approach to transform diopside bulk material into a 2D.

It should be stressed that most of the obtained 2D materials come from layered materials, which poses some limitations. The first 2D material obtained from non van der Waals solids was the hematene[22] (obtained from 3D hematite). Other structures followed the same experimental approach (liquid exfoliation), such as ilmenite[23], chromite[24], etc. However, up to now, no 2D silicates were experimentally realized. In this paper, we present the first method to create 2D d-silicates from natural ores; named Ore Exfoliated Silicates (OES). We have utilized various spectroscopy (Raman, XPS, FTIR, etc) and microscopy (AFM, SEM, and TEM) tools to characterize the obtained exfoliated 2D sheets: 2D *d*-silicates (two-dimensional diopside (*d*) Silicates). DFT simulations were used to gain further insight into the structural characteristic and electronic features. The method is easy and scalable and can be used for obtaining other silicates.

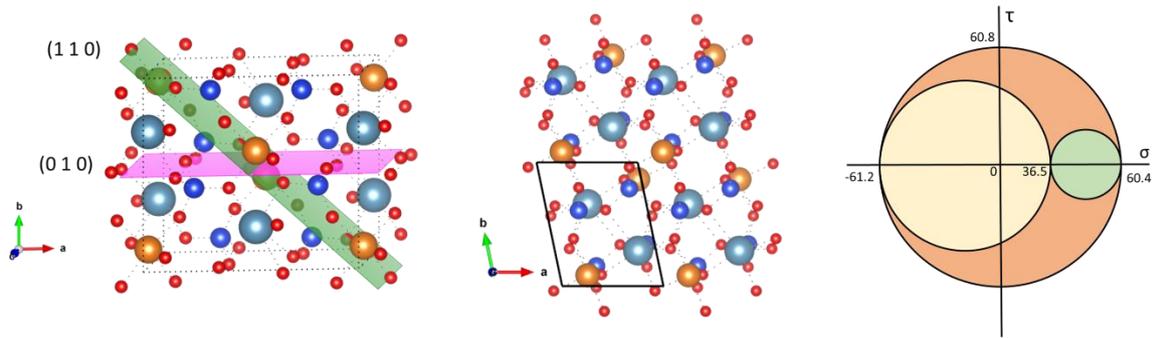

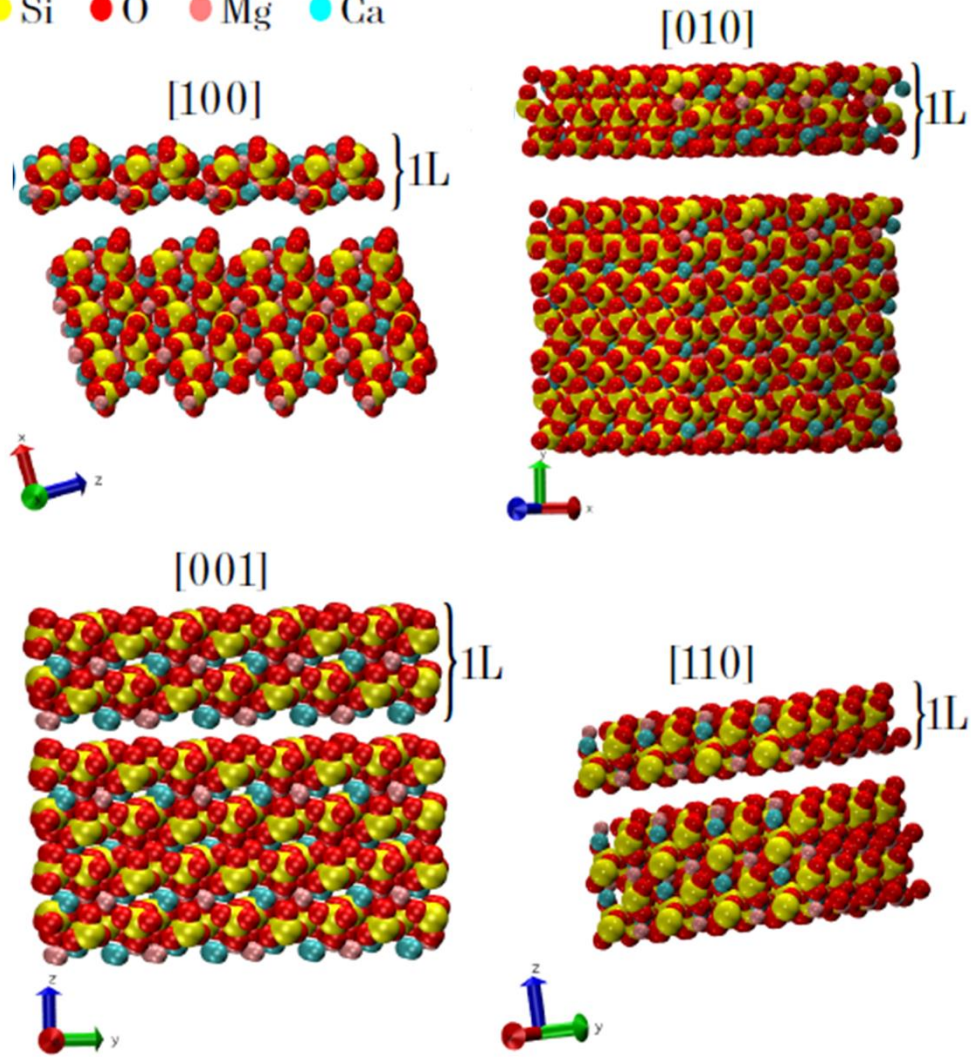

*Figure 1*: *(Top-left) Diopside bulk highlighted with the possible exfoliation planes: (0 1 0) (pink color) and (1 1 0) (green color); (Top-middle) the (1 1 0) exfoliated plane; atom colors representation: Ca in cyan-blue, O in red, Si in blue, and Mg in gold, (Top-right) Represents the pure shear condition for the Diopside in form of Mohr Circle for 3D stress tensor and;*

*(bottom panel) Optimized atom positions, lattice vectors and angles for diopside exfoliation along [100], [010], [001], and [110] crystal directions.*

## 2. Experimental section

The bulk diopside mineral was obtained from the localities of Kimberley - South Africa as crystal. The bulk size crystals were first ground in a mortar pestle to make it a fine powder, this powder is then taken in ~0.05 grams in 50 mL IPA and probe sonicated for 5 hours. The temperature of the solvent was kept below 35°C. The temperature rise was controlled by giving 15-20 minutes intervals after every 30 minutes of sonication. The suspended stable solution was separated and kept as the actual 2D *d*-silicate samples and used in further experiments. The synthesis process is schematically depicted in **Figure S1**.

To fabricate the OES-based device for nanoscale energy harvesting, a piece of filter paper (5.4 x 1.2 $cm^2$) was immersed into the 2D d-silicate solution and kept until the solution dries with the surface fully coated with the exfoliated samples. The coated substrate is then fabricated into a device by putting Aluminium (Al) electrodes, on both sides, and the lids were applied to connect with the digital oscilloscope (Tektronics) during electrical measurements. The fabrication design is schematically shown in **Figure S2 (a)**.

An electron probe micro-analyzer (EPMA) study was carried out at Rice University, Department of Earth, Environment and Planetary Science using a JEOL JXA 8530F Hyperprobe to study the actual composition. X-ray diffraction (XRD) analyses were performed using *Bruker D8 Advance with Lynx eye detector* with Cu Kα radiation of wavelength λ= 0.15406 nm in the 2θ range of 10 to 90˚. The morphology of the samples was studied using a Scanning electron microscope (SEM, Sigma 300) and a thin layer of gold was sputtered on samples to avoid charging artifacts. The atomic orientation and exfoliated sample imaging were done using a High Resolution-Transmission electron microscope (HRTEM, Titan Themis 300 kV). The thickness of the exfoliated 2D samples was characterized with an atomic force microscope (Agilent Technologies, 5500, cantilever length of 100 microns, and tapping frequency of 150 kHz). Raman spectra were obtained with *T64000*, *Jobin Yvon Horiba, France make spectrometer* with a 532 nm excitation wavelength. Fourier Transform-Infrared Spectrometer (FTIR, *Thermo fisher scientific,* NICOLET 6700) was used for qualitative analysis of inorganic components in transmittance mode. A UV-Vis spectrophotometer (Analytical) is used for the absorbance and bandgap analysis of the materials.

## *Computational details*

We carried out density functional theory (DFT) simulations for the bulk and 2D *d*-silicate (110). Firstly, the (110) (the most commonly observed in the experiments) surface was cleaved to create a 2D structure from the bulk one. As standard for studying 2D structures, we use a vacuum buffer region of 15 Angstroms. The created unit cell (slab) was fully optimized (atomic positions and cell parameters). We used the CASTEP Code[25] with the following convergence criteria: energy $5 \times 10^{-5}$ eV/atom, maximum. force value of 0.1 eV/Å, maximum stress value of 0.2 GPa, and maximum atomic displacements of 0.005 Å. Generalized gradient (GGA) functionals–WC as exchange-correlation functional and On-the-fly (OTFG) ultrasoft pseudopotentials were used for the plane wave (PW) basis set to treat the valence electrons[26,27]. The chosen algorithm for performing this task was Broyden-Fletcher-Goldfarb-Shanno (BFGS) minimizer to obtain the lowest energy structure at a fixed external stress[28]. The Monkhorst-pack grid used was of 4x4x1 k-points to sample the Brillouin zone and the energy cut-off was set to 381 eV[16]. The self-consistent field (SCF) was allowed to converge to $10^{-6}$ eV/atom.

Once the structures were optimized, we estimated the bond energy value $E_{bond}$ for the layers [100], [010], [001], and [110] by the expression:

$$E_{bond} = (E_{layer} + E_{substrate} - E_{layer/substrate})/A^2, \qquad (1)$$

where the $E_{Layer}$ and $E_{substrate}$ are the energies of an isolated layer and substrate, $E_{layer/substrate}$ is the energy of the layer/substrate complex corresponding to the chemisorption system, and A is the contact area between layer and substrate[15].

## 3. Results and discussion

**Figure 1** shows the crystal structure of diopside bulk and some possible exfoliation planes and its corresponding calculated Mohr Circle at the top. At right panel, we have the slabs for each exfoliation plane along the [100], [010], [001], and [110] crystal directions. We use the expression (1) to obtain the bond energy value and we assume that the substrate is the same diopside composite from the layer.

For the cases shown in Figure 1, we obtained the following values for exfoliation energy: 1.4, 0.4, 7.7, and 3.1 kcal/molÅ$^2$, corresponding to the [100], [010], [001], and [110], respectively. We observe that for all cases the exfoliation energy is positive, which means that the process is exothermic, in accordance with other silicates systems[15]. The smallest value for exfoliation energy was observed for the [010] crystal direction indicating that exfoliation should be easier for this direction in comparison with the other ones.

The composition of naturally obtained bulk diopside was initially studied by EPMA analysis (**Figure S3**) to quantify the elemental weight percent, including the impurity atoms, and also in the form of oxide compounds. In **Figure S3(a)**, each element present in the natural mineral diopside is normalized against six oxygen atoms and represented in a bar graph. The quantification shows that the presence of basic elements of the ideal diopside compound ($CaMgSi_2O_6$) such as Si, Mg, and Ca has a solid solution with the impurity atoms such as Na, Fe, Cr, Al, Ti, Mn, and Ni in minor concentrations. The oxide phases such as $SiO_2$, $CaO$, $MgO$, $FeO$, $Cr_2O_3$, $Na_2O$, and so on were represented in **Figure S3(b)**, where a higher contribution of $SiO_2$, CaO, and MgO were observed. Upon calculation of each element's contribution through EPMA, the final composition formula is represented as:

$(Ca_{0.879}Na_{0.073}Mg_{0.037}Mn_{0.002})_{\Sigma=0.991}(Mg_{0.879}Fe^{2+}_{0.062}Cr_{0.037}Al_{0.015}Ti_{0.006}Ni_{0.001})_{\Sigma=1.00}Si_{2.006}O_6$

The physical characteristics of this mineral show a greenish color appearance (**Figure 2(a)**). The process of bulk crystal exfoliation into thin flakes is represented in **Figure S1,** and the detailed procedure is mentioned in the experimental section. To understand the difference in terms of morphology before and after exfoliation, the crushed bulk sample was first imaged with SEM, where a large ~6-micron size particle is observed, as shown in **Figure 2(b)**. Following exfoliation, the flake lateral size was 600 nm, which is more than ten times smaller in size (**Figure 2(c)**). Figure 2(b) shows bright-field low magnification TEM images of separating layers caused by exfoliation. 3 layers of exfoliated sheets are shown in **Figure 2(d)**. The monolayer of the diopside crystal can be identified in **Figure 2(e)**. The high-resolution 3D plot of the TEM image shown in **Figure 2(f)** depicts (220) projection of planes, where several atomic defects and dislocation of atoms were observed. These defects in the crystal will lead to dangling bonds on the surface of 2D *d*-silicate-like materials upon exfoliation and create an abundance of active sites[17]. **Figure 2(g)** shows the (0 2 0) plane, where dislocated atoms were observed along with several atomic vacancies in the plane. The projection of the crystal pattern is shown in the schematic (**Figure 2(h)**), crystal orientation along (0 2 0) also shows separate layers of Ca, Mg, and Si atoms. The selected area electron diffraction (SAED) pattern shown in **Figure 2(i)** shows the single-crystalline nature of the sample where dot patterns were observed for (1 1 -1), (1 -1 1), and (2 0 0) directions. To determine the flake thickness, Atomic Force Microscopy (AFM) analysis was carried out and shown in **Figure 2(j)**. A major contribution (>45%) of 0.55-0.6 nm flake thickness is observed which matches with monolayer thickness predicted along [100], along with a minor contribution (5-15%) from higher thickness (1-2 nm) flakes, as shown in the histogram (**Figure 2(k)**). The lateral length of the flakes was observed to be in the range of 180 nm to 370 nm, from AFM analysis (**Figure 2(l)**). However, ~600 nm flake size was also observed in TEM analysis (**Figure 2(b)**), which suggests the presence of large lateral size but with very low thickness of the exfoliated sheets, therefore calling it a 2D structure is justified.

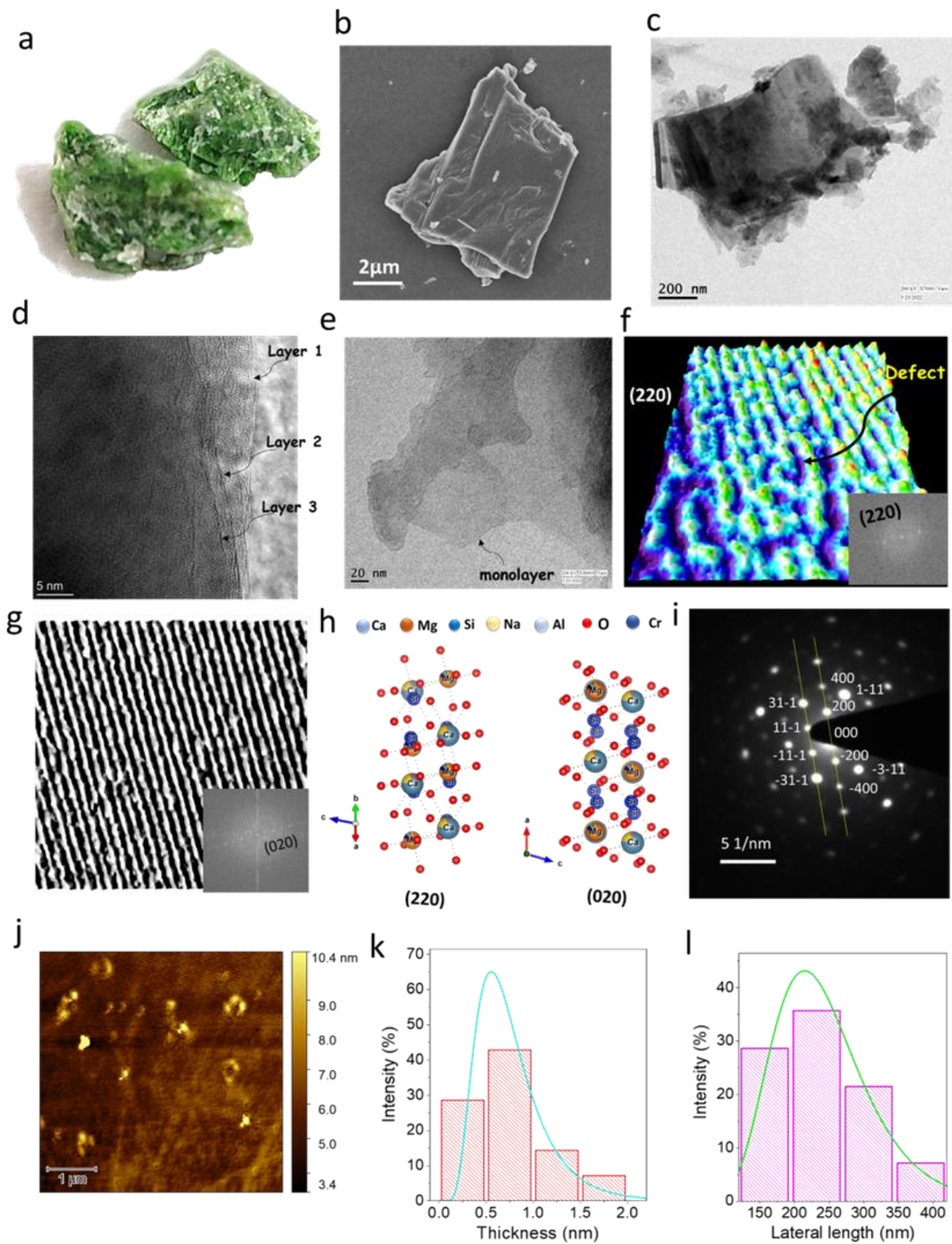

***Figure 2***: *(a) Digital image of bulk diopside sample, (b) SEM image of the crushed bulk sample, (c) Bright-field low magnification TEM image of exfoliated diopside flake, (d) High-resolution image of exfoliated sheets with 3 layers, (e) Magnified TEM image showing a monolayer of*

*diopside after exfoliation, (f) 3D view of (220) projection of diopside crystal and the corresponding FFT pattern in the inset, and (g) high-resolution TEM image showing (020) plane with the FFT pattern in the inset, (h) α and β patterns showing crystal orientation along (220) and (020) directions, (i) SAED-TEM pattern of diopside, (j) AFM image analysis of the exfoliated sample, (k) histogram of flake thickness profile and (l) histogram of the lateral length of exfoliated flakes from AFM image.*

The XRD pattern of bulk and 2D *d*-silicate sample (**Figure 3(a)**) shows the presence of Fe and Al ($Al_{0.17}Ca_{0.83}Cr_{0.05}Fe_{0.07}Mg_{0.86}Na_{0.1}O_6Si_{1.91}Ti_{0.01}$). The bulk and 2D crystal structures are monoclinic and have a C12/c1 space group, with lattice parameters a = 0.98 nm, b = 0.9 nm, and c = 0.528 nm. The crystal, even after exfoliation for several hours, still shows its polycrystalline nature, however, several plane intensities have changed upon exfoliating the bulk. The maximum intensity of the exfoliated plane here is at (220) or (110). This can be confirmed by the observed HRTEM pattern in **Figure 2(f)**. A major decrease is observed in (0 2 -1), (3 1 -1), (4 0 0), (1 1 2), (4 0 -2), (5 3 2), (3 5 -2), and (7 1 -1) planes and they have almost disappeared. This change indicates a distortion in the monoclinic structure in these directions. This is a typical finding during exfoliation, where the breaking of bonds causes the destruction of planes and structural flaws. The 3D crystal structure of diopside is monoclinic with symmetry group C2/C, where the Ca atom sits in an octahedral position and all other atoms Mg, Si, and O favors a tetrahedral position. The transformation of 3D diopside to 2D did not change anything structurally except the atoms sitting at the top site, because of the truncation of the edge atoms' bonding. The 2D structure re-assembles by the association of new bonds and closely resembles the triclinic structure with the P-1 symmetry group. This induces the asymmetric charge distribution needed for energy harvesting.

A Raman spectroscopy study was also conducted to observe the vibrational states of the exfoliated samples (**Figure 3(b)**). Raman peaks of lattice vibrations or phonons in 2D materials exhibit several distinctive features, such as peak position, FWHM, intensity, and peak shifts, providing valuable information about phonons, electron-phonon coupling, and electronic states[29]. It is used to probe the electronic band structure and other electrical properties in addition to characterizations of the structural properties, like the number of layers, defect densities, strain, etc[30]. At 1008 cm$^{-1}$ ($F_{2g}$), a significant stretching of the Ca-O bond is seen. Minor peaks at 970-945 cm$^{-1}$ signify stretching of Si-O bond along with Ca-O bond vibration at 865 cm$^{-1}$. A significant Si-O(-Ca) bond vibration was detected at 665 cm$^{-1}$ ($F_{2g}$). Bending vibrations in the [MgO$_4$]$^{6-}$ tetrahedron and Si-O(-Ca) were seen at 392 cm$^{-1}$ ($F_{2g}$). The peak at 368 cm$^{-1}$ represents Si-O bending, and 359 cm$^{-1}$ suggests Si-O(-Mg) vibrations. Si-O(-Mg) bond vibrations were also seen at 323 cm$^{-1}$ ($F_{2g}$), while at lower frequencies of 302 cm$^{-1}$ and 213 cm$^{-1}$ represent Ca-O and Si-O(-Ca) symmetric stretching respectively. FTIR analysis was also carried out to verify the bond vibrations, as shown in **Figure 3(c)**. An amount of energy corresponding to the spectrum's infrared region is typically needed for vibrational transitions. In this 2D *d*-silicate sample, similar to the Raman spectra observation, mostly Si-O(-Ca) bonds were observed as prominent and lower frequency vibration from Mg$^{2+}$ tetrahedron. However, the bulk sample shows the major presence of O-Mg-O, O-Ca-O bond vibrations, and minor Si-O vibrations.

Light absorbance in the ultraviolet region is revealed by UV-vis spectroscopy analysis of 2D material's optical properties, which suggests that high energy is required for the transition of electrons from the valence to the conduction band. The exfoliated 2D material has a bandgap of about 4 eV, making it a wide bandgap material (**Figure 3(d)**). These wide-bandgap materials are suitable for high-temperature applications. The following atomic notation characterizes the valence states of the constituent atoms in the diopside: Mg (*2p, 3s*), Ca (*3p,4s*), Si (*3s,3p*), and O (*2s,2p*). **Figures 3(e) and 3(f)** show the band structure of the 3D structure with a direct band gap value of 3.572 eV and the 2D structure with a direct band gap value of 4.176 eV, which are in good agreement with experimental ones. In comparison to 2D band structures, we observe upward shifting of conduction bands by ~2eV because of the majority from Si '*3p*' states and minor from Ca *4s, 3p,* and Mg *2p, 3s* states in the density of states plot, contributing to anti-bonding orbitals, π*$_{p-p}$. Near the Fermi region, O '*2p*' states with Mg '*2p*' states remain predominant and in bonding states near ~-16eV, σ- bonding of O '*2s*' states with Mg '*2p*' states is predominant.

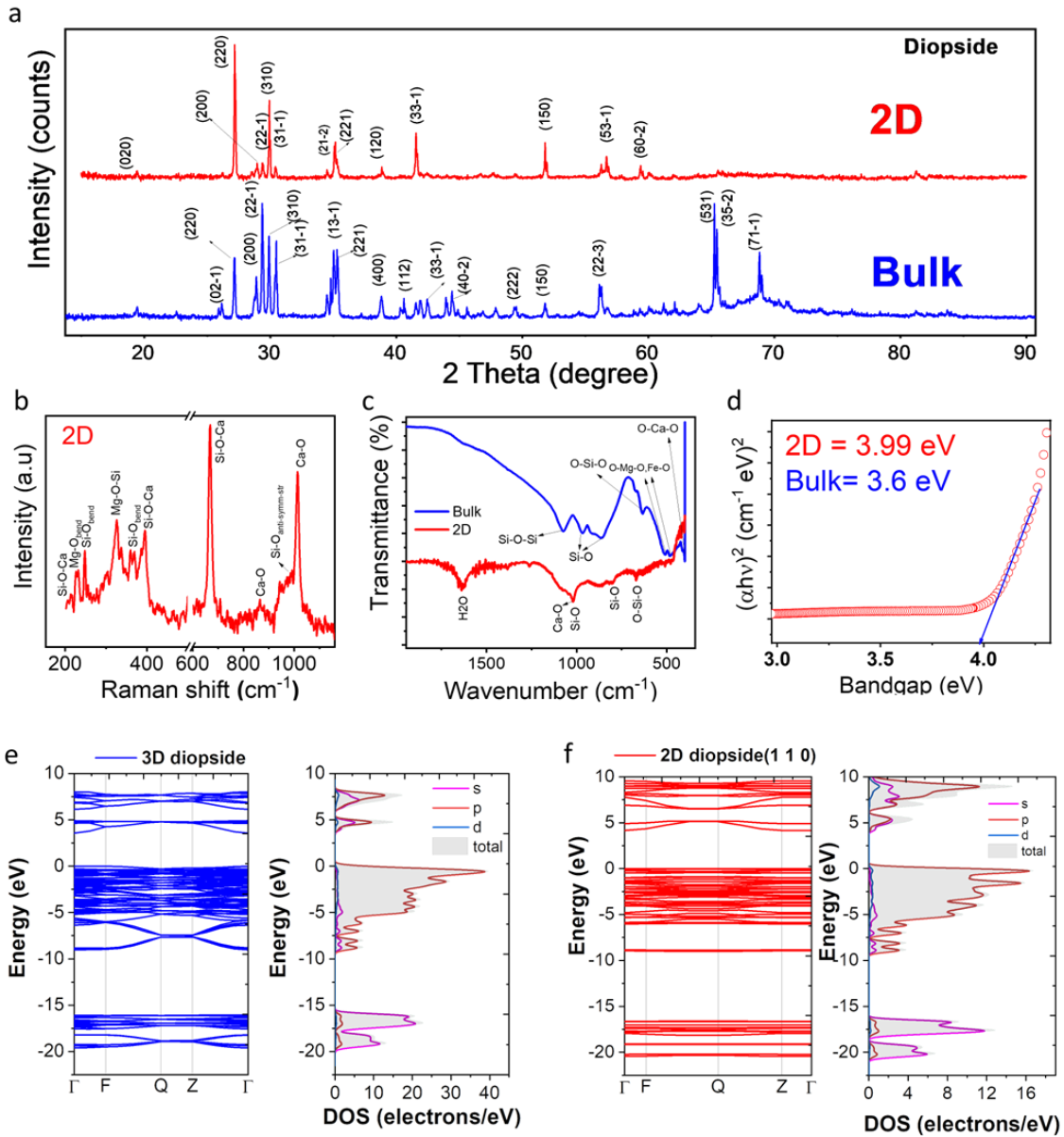

*Figure 3*: *(a) XRD pattern of bulk and exfoliated diopside sample, (b) Raman spectra of exfoliated diopside sample, (c) FTIR spectra of bulk and 2D d-silicate, (d) Absorbance and*

*band gap plot of the exfoliated sample (experimental), the band structure of diopside for (e) Bulk and (f) 2D.*

To confirm the presence of elements and their quantification in terms of atomic percentage, Energy Dispersive X-ray Spectroscopy (EDX) was carried out. **Figure S4** shows EDX spectra of exfoliated diopside samples, where along with Ca, Mg, Si, and O, elements such as Na, Al, Ti, Cr, and Fe are also present. These elements are present in solid solution with the idle composition of diopside. The exposed composition suggests a major presence of iron (Fe), sodium (Na), and aluminum (Al) apart from $CaMgSi_2O_6$. The inset of **Figure S4** shows the calculated atomic percent from the EDAX spectra.

As observed from XRD spectra, some reorientation of planes can also create defects as observed in TEM images. **Figure 4(a)** shows similar elements (Ca, Al, Mg, Si, O) present at the surface upon X-ray photoelectron spectroscopy analysis. A major oxide contribution was observed in the O 1s state with a binding energy of 532 eV. The deconvolution of O1s spectra revealed that 30% of the oxide groups were in chemisorbed oxygen states, with the remainder in metal-oxide bonding states (**Figure 4(b)**). Similarly, the calcium (Ca) exists here in the 2p state at 348 eV where the Ca $2p_{3/2}$ state has significant contribution bonding with either OH, F, or Cl (**Figure 4(c)**). Along side, Ca 2p ½ state is also present at 357 eV with a minor satellite peak at 336 eV. In **Figure 4(d)**, the Mg KLL state suggests a major contribution from the Mg-O state and a minor contribution from metal-metal bondings. These KLL states represent electron energy ejected from the atom as a result of the L shell electron filling the K shell and the L shell electron being ejected at the same time. In **Figures 4(e)** and **4(f)**, silicon 2s and 2p states were represented with a binding energy of about 154 eV and 103 eV, respectively. Si 2s spectra show an almost similar contribution from both oxide and silicate states.

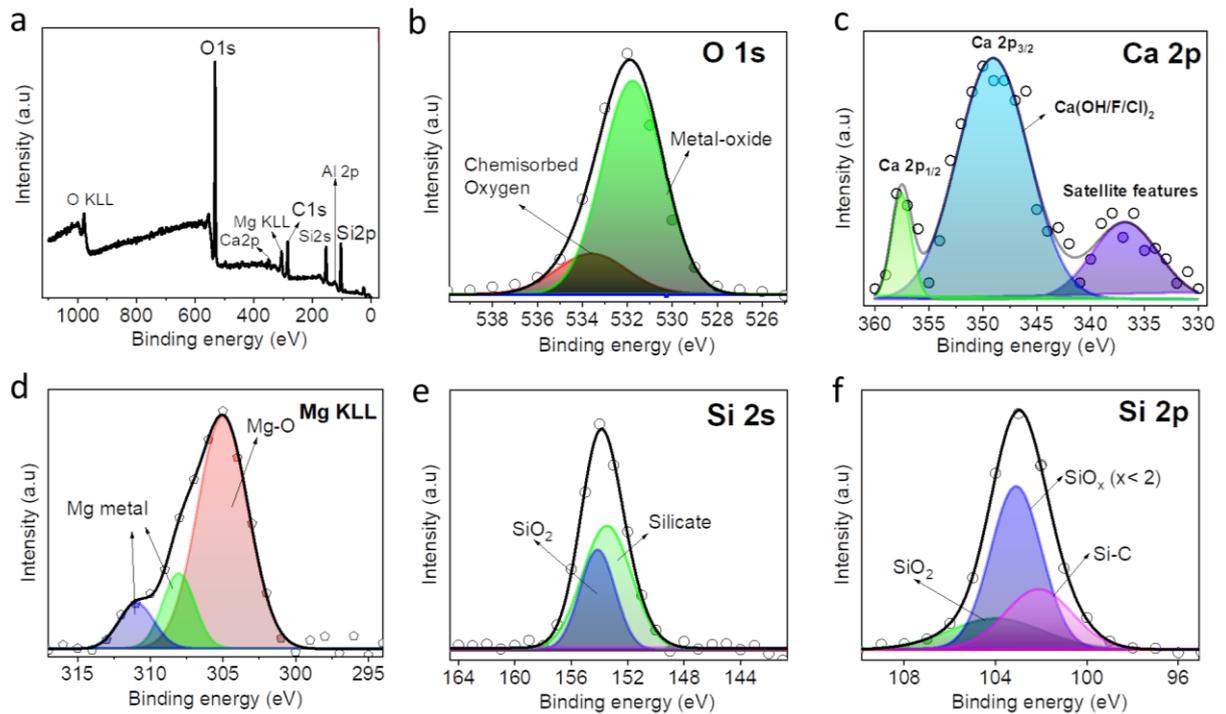

*Figure 4*: XPS analysis spectra showing (a) full surface scan showing all the elements present at the surface, (b) O1s spectra, (c) Ca 2p spectra, (d) Mg KLL spectra, (e) Si 2s spectra, and (f) Si 2p spectra.

After characterizing and confirming the 2D nature of *d*-silicates, we have investigated the electrical output performance of 2D *d*-silicates-based flexible energy harvesting device. The electrical conductance, impedance, capacitance, and dissipation factors were also determined at 307K room temperature and different frequencies to understand the device characteristics. Capacitance voltage measurements were made at zero bias (**Figure 5(a)**) to study the frequency range and self-resonant frequency of the fabricated device. Capacitance changed from 55.4 pF to 44 pF. It is observed that there is a sudden drop in the capacitance value up to 10kHz and above that the capacitance (~44pF) almost remains constant. This shows that the fabricated device has a resonant frequency in between these ranges of frequencies. As expected, the decrease in capacitance has reduced the overall impedance of the device above 10kHz (**Figure 5 (b)**) and increased the conductance of the device from 4.87 nS to 830 nS (**Figure 5(b)**). The dissipation factor is associated with self-heating under resonant conditions. and went from 0.14 to 0.002 from 0.1kHz to 1MHz frequency (**Figure 5(c)**). It is observed that the dissipation factor dropped to a very low value from 100 kHz and above. This shows that this device can be used for resonator applications due to its lower dissipation factor.

In **Figure 5(d)**, it can be observed that various force/stress values have a different impact on the device's electrical output. The generated voltage output at the positive half cycle varied from ~2V to ~5V for 0.98N to 5.88 N under stress application. Here, we have rectified the electrical response at 0.98 N in the positive cycle, with an output of ~2V as shown in **Figure 5(e)**. The internal resistance of the fabricated device was estimated to be around 2kΩ with a voltage output of ~0.258 V (**Figure 5(f)**). From this, the device's power was calculated to be about 33.28 µW and the power density is about 0.0513 W/m$^2$. **Figure 5 (g)** shows a peak-to-peak voltage output bar graph when different load values are applied. Similarly, the performance of the device with 0.98N force application at different temperatures applied to the device can be understood through the peak-to-peak voltage plot in **Figure S5.** This device was mounted on the human body to study the real-time application and performance of the exfoliated 2D sample. **Figure S6** shows different electrical outputs generated due to stress application techniques, either through tapping, bending, mounting, and movement of the body, and so on. The device when mounted on a finger moved vertically for several cycles generating a voltage up to ~400 mV and responds like a robotic finger (supplementary video V1). The mounted device acts as a motion sensor and gives signals accordingly.

To better understand the energy harvesting mechanism, we studied the Electric field gradient (EFG) response of the material. Interestingly, density at the nucleus is zero for all the atoms which indicate that the atom nuclei are in the nodal region, and the probability of finding an electron would be at the points in the wave where concavity changes. **Table 1** (Supplementary file) summarizes the asymmetry factor, $\eta$, and electrostatic potential V along the x, y, and z directions. Si (top and bottom sites), Mg atom sites, and surrounding O atom sites have the most asymmetry factor of ~0.6, but low electrostatic potential ~ $3.5 \times 10^{21}$ V/m$^2$. This situation leads to smaller charges and favors shorter distances. While Si (middle layer) atom sites and surrounding O atom sites have higher charges and relative distance to oxygen atoms is greater than the top and bottom atom sites. It has the least asymmetry factor ~ 0.2 but the high electrostatic potential ~ $7.5 \times 10^{21}$ V/m$^2$ along the z direction. This phenomenon can be visualized by plotting dipole in x-y plane atomic coordinates and z-view projection of atom sites, represented in **Figure 5(h).**

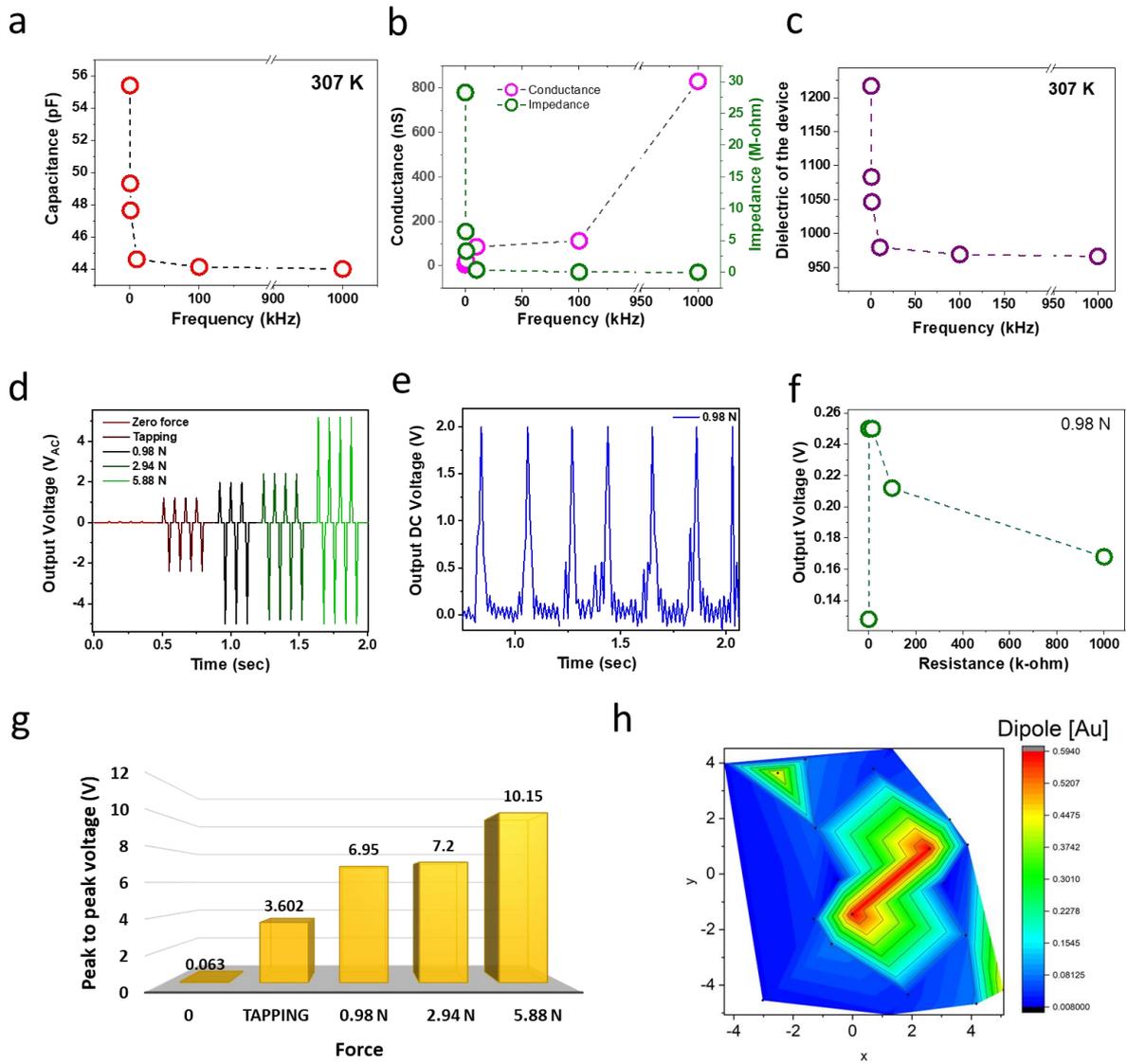

*Figure 5*: *(a) Capacitance at 307K for different frequency values, (b) Conductance and Impedance plot at different frequency values at 307K, and (c) Dissipation or loss factor of the fabricated 2D d-silicate device at 307K, (d) Voltage output signal depending on different force applications in AC, (e) Rectified DC voltage signals upon bridge rectifier implementation, (f) Resistance versus voltage plot with 0.98 N force for internal resistance calculation, (g) Peak to peak voltage output comparison bar graph for different force application, and (h) Charge density distribution contour plot projected in the x-y plane.*

Up to 8.8N force we have an 1.5% increment in asymmetry factor that includes a 0.2% reduction in Mg-O bond length at the edge sites, followed by a 0.1% reduction change in Ca-O and Si-O edge sites. Bond length fluctuations are connected to the response of the material in reaction to the applied external force. The atomic strain results in charge asymmetry and local electric field gradient are further enhanced, which results in the flexoelectric coefficient maximum of 21.318 nC/m along the z-direction, whereas 2.62 nC/m along the x-direction, and 8.68 nC/m along the y-one. In summary, the above analysis shows a great sensitivity of the electric field gradient with respect to the charge distributions and bond length change.

**Conclusion**

In summary, we have demonstrated a simple and easily scalable technique of liquid phase exfoliation method to synthesize the first 2D *d*-silicate from a non-layered natural silicate mineral called diopside. X-ray, TEM, HRTEM, Raman, and FTIR techniques were used to characterize and confirm the existence of 2D structures. A device consisting of a thin coating of 2D *d*-silicate was fabricated. The energy harvesting device results in the generation of around 10V under a compressive load of 5.88 N. The structural stability and energy harvesting mechanism were further validated by DFT simulations. The current approach is completely general and can be utilized for large-scale synthesis of 2D silicates and their derivatives, whose large-scale syntheses have been elusive.

**Competing Interest Statement**: The authors have declared no competing interest.

# Supporting information

**Synthesis procedure**

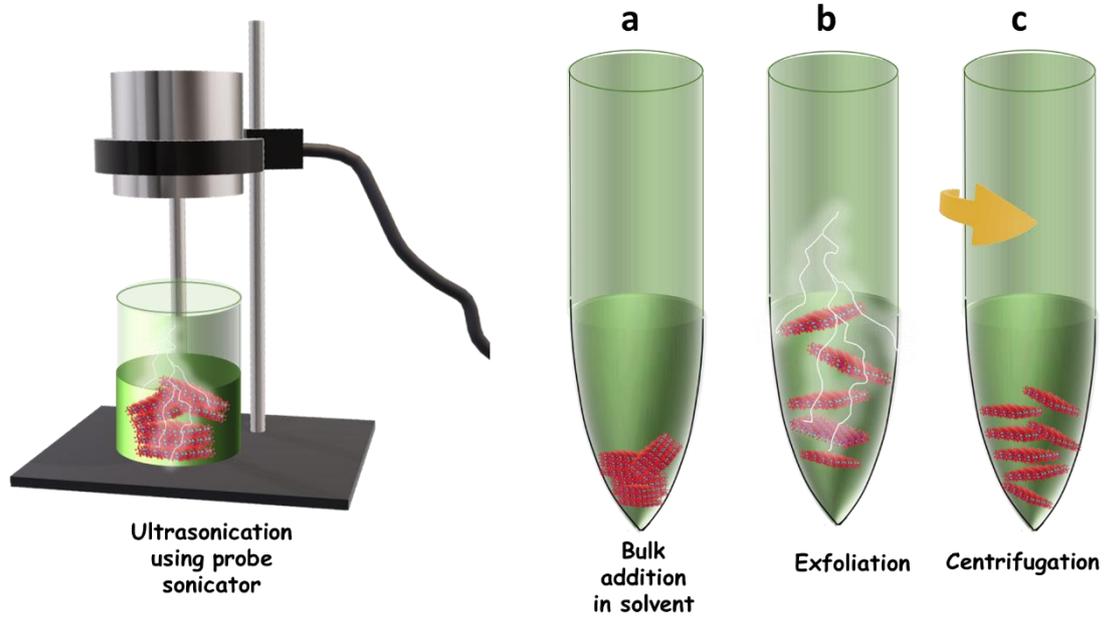

***Figure S1****: Synthesis of 2D d-silicate via liquid phase exfoliation.*

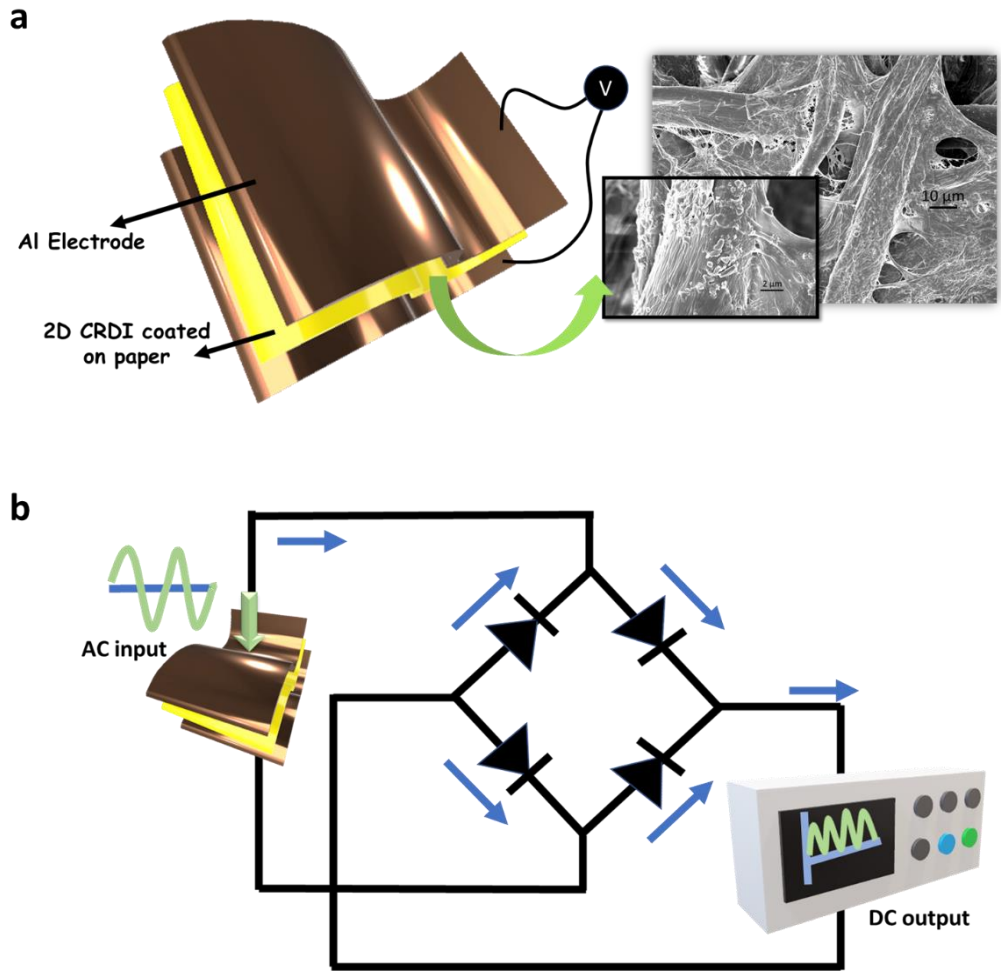

*Figure S2*: *Schematic of (a) fabrication design of flexible energy harvesting device with 2D d-silicate and SEM image of cellulose interaction with 2D d-silicate flakes, and (b) Bridge-rectifier circuit for converting AC voltage into pulsating DC voltage for use in power supplies.*

**Response measurement conditions:**

Here the measurements were carried out at room temperature of about 307K, the force values were varied from 0 to 5.88N. The device was kept on an insulated surface (paper sheet) and force was applied from one side of the device. The voltage output readings were recorded with a connected oscilloscope system. The noise value upon no load application was around 0.2V due to environmental factors. The signals are alternating positive and negative cycle sinusoidal waves due to the capacitor configuration. For application in DC devices where direct charging is required, a rectified electrical signal is needed. Mostly the AC to DC signal is achieved by employing a bridge rectifier circuit (as shown in Figure S2 (b)) where the signal can be rectified either to the positive or negative side.

**Electron probe micro-analysis (EPMA)**

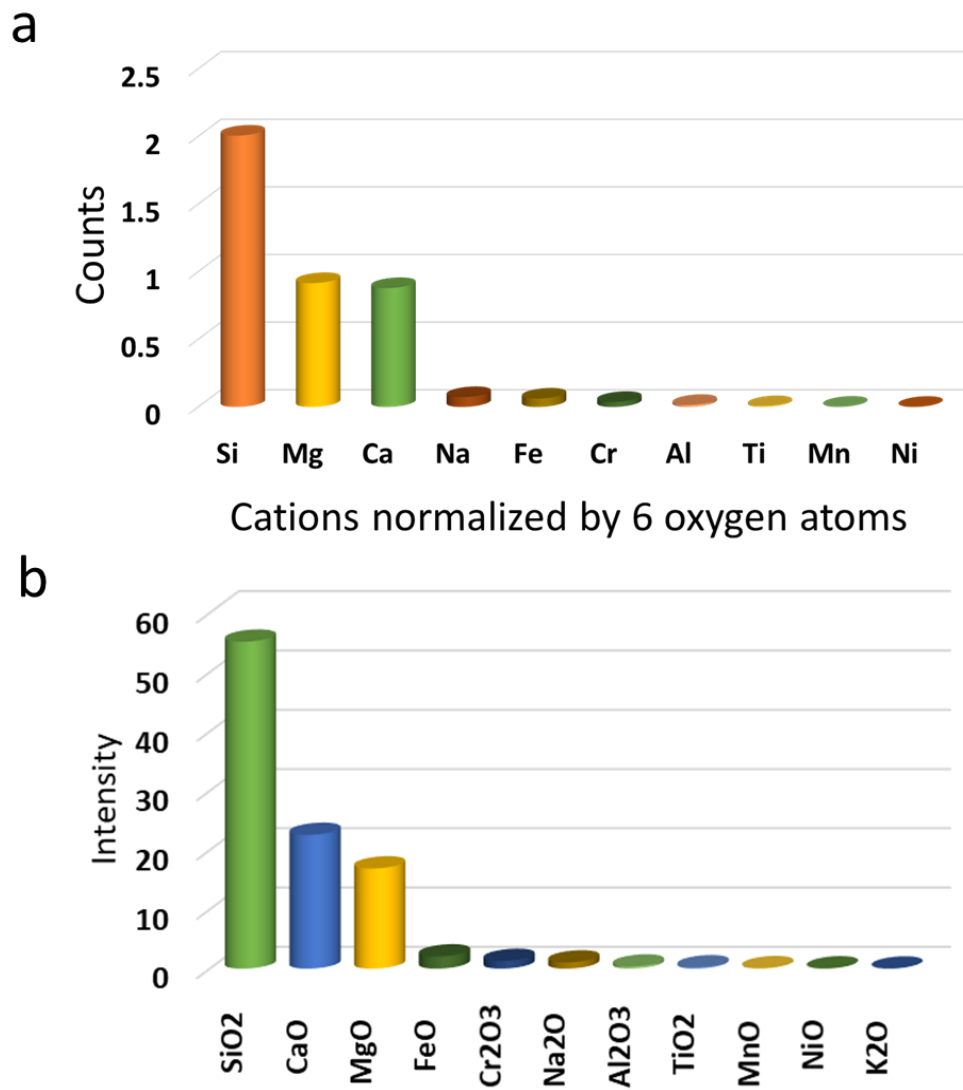

*Figure S3: (a) EPMA study of cations normalized by 6 oxygen atoms, and (b) EPMA quantification of oxide compounds in bulk diopside.*

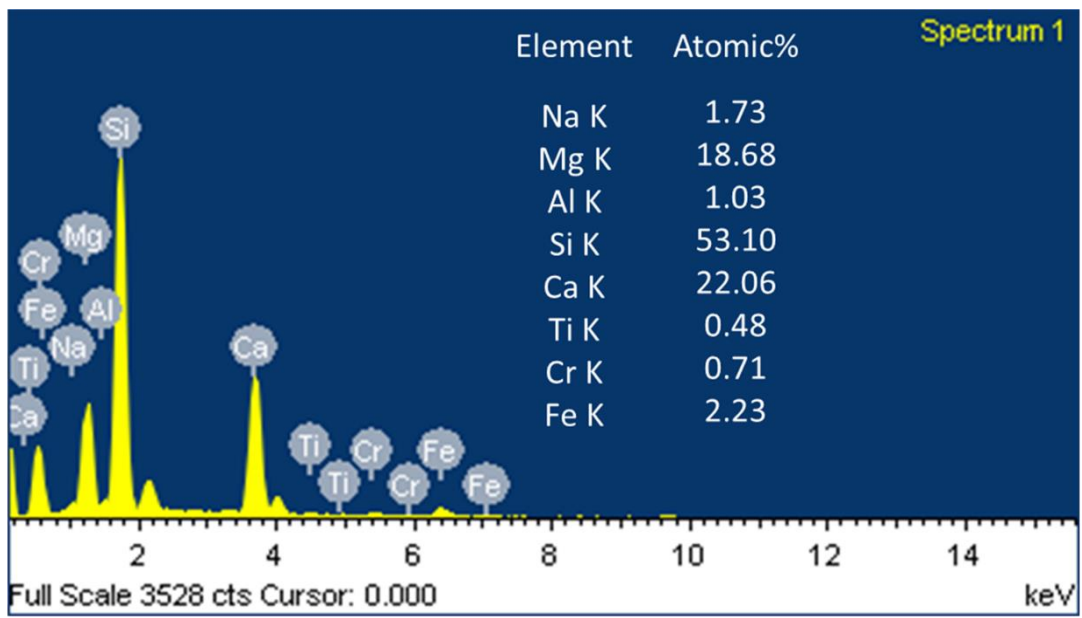

*Figure S4*: *EDX spectroscopy analysis showing atomic % of each element present.*

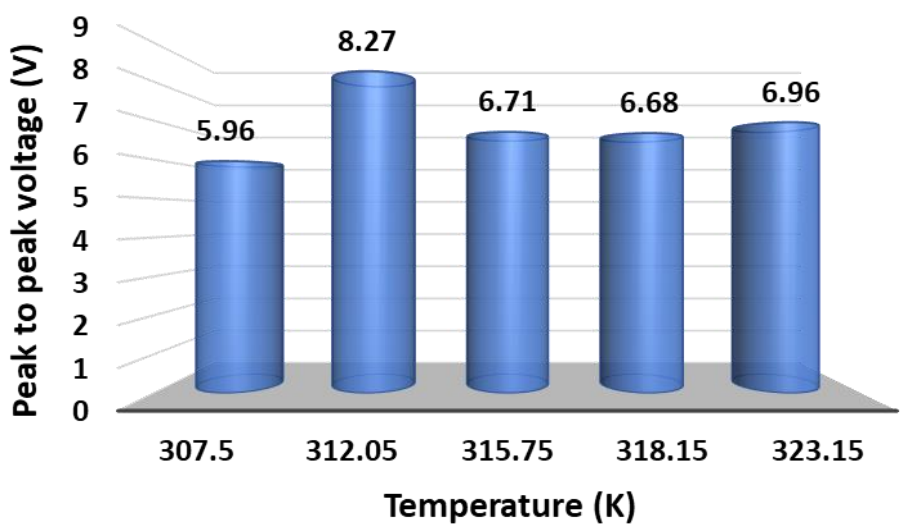

*Figure S5*: *Bar graph showing net peak to peak voltage output at various temperatures(K) with 0.98 N force.*

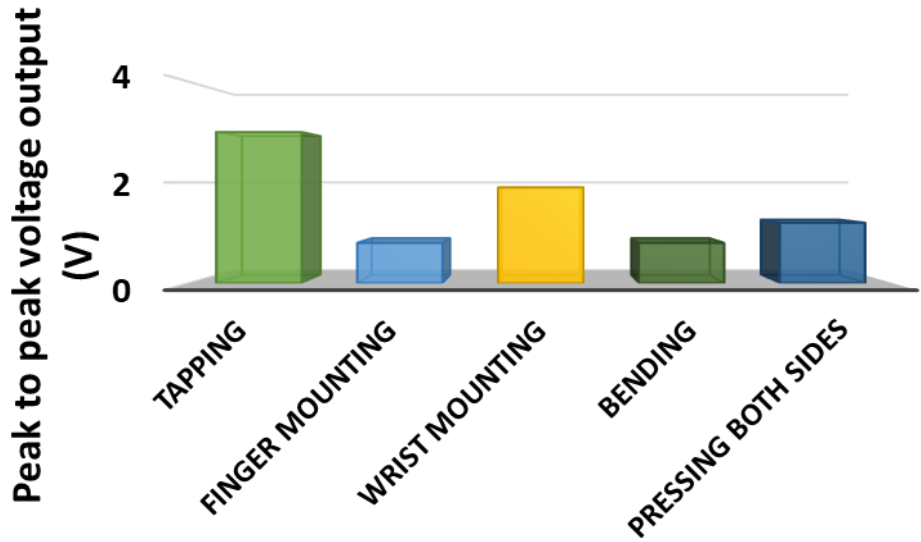

*Figure S6*: Bar graph showing peak to peak voltage output for different physiological movement with the 2D *d*-silicate device.

Table-1: Anisotropy calculation summary for 2D *d*-silicate (1 au = 0.97 x $10^{22}$ V/m$^2$).

| Atoms | $V_{xx}$ [au] | $V_{yy}$ [au] | $V_{zz}$ [au] | $\eta$ | Mulliken charges (e) |
|---|---|---|---|---|---|
| O | -0.358 | 0.2898 | 0.0682 | 0.2057 | -1.23 |
| Si | 0.247 | 0.0228 | -0.2698 | 0.6229 | 1.97 |
| O | -0.3281 | -0.1963 | 0.5244 | 0.3163 | -1.21 |
| O | 0.5102 | -0.3348 | -0.1754 | 0.2926 | -1.16 |
| Si | 0.3225 | -0.5548 | 0.2323 | 0.2881 | 1.92 |
| O | -0.3815 | 0.6316 | -0.2501 | 0.2471 | -1.10 |
| O | -0.1479 | 0.5755 | -0.4276 | 0.5522 | -1.19 |
| O | 0.548 | -0.2804 | -0.2676 | 0.0226 | -1.16 |
| Si | 0.2484 | 0.0185 | -0.2669 | 0.632 | 1.97 |
| O | -0.3577 | 0.297 | 0.0607 | 0.2046 | -1.23 |
| O | -0.3249 | -0.2043 | 0.5292 | 0.296 | -1.21 |
| O | 0.5103 | -0.3371 | -0.1732 | 0.2943 | -1.17 |
| Si | 0.3207 | -0.5515 | 0.2307 | 0.2874 | 1.92 |
| O | -0.3854 | 0.6319 | -0.2466 | 0.2472 | -1.10 |
| O | -0.1501 | 0.5762 | -0.4261 | 0.5489 | -1.19 |
| **O**  | 0.5486  | -0.281  | -0.2676 | 0.0177 | -1.16 |
| **Mg** | 0.0246  | -0.1701 | 0.1455  | 0.5555 | 1.45  |
| **Ca** | -0.1867 | -0.2305 | 0.4172  | 0.3855 | 1.71  |
| **Mg** | 0.02    | -0.165  | 0.1451  | 0.5973 | 1.45  |
| **Ca** | -0.1812 | -0.2363 | 0.4175  | 0.3664 | 1.71  |